\documentstyle[aps,prl,twocolumn,epsf]{revtex}

\begin{document}
\title{ Group Theory Approach to Band Structure: Scarf
  and Lam\'e Hamiltonians}

\author{Hui Li and Dimitri Kusnezov
     \footnote{E--mail: dimitri@nst4.physics.yale.edu} }

\address{ Center for Theoretical Physics,
              Sloane Physics Laboratory,
             Yale University, New Haven, CT 06520-8120 \\}

\date{November 30, 1998}

\maketitle
\vspace{3cm}

\begin{abstract}
The group theoretical treatment of bound and scattering
state problems is extended to include band structure. We show
that one can realize Hamiltonians with periodic potentials as
dynamical symmetries, where representation theory provides
analytic solutions, or which can be treated with more general
spectrum generating algebraic methods. We find dynamical
symmetries for which we derive the transfer matrices and
dispersion relations. Both compact and non-compact groups are
found to play a role.
\end{abstract}

\pacs{PACS numbers: 03.65.Fd, 02.20.-a, 02.20.Sv, 11.30.-j}

%

\narrowtext

The application of group theoretical techniques to physical problems has a long
and fruitful history. From the early works on the low-lying spectrum of
hadrons\cite{early}, to the more recent studies of nuclear\cite{nuc} and 
molecular\cite{mol} many-body problems, the theory of Lie groups and algebras
has become important not only in explaining the behavior of various physical
systems, but also in constructing new physical theories. Problems of interest
can usually be approached in this manner when a spectrum generating algebra
(SGA) can be  identified. A SGA exists when the Hamiltonian $H$ can be
expressed in terms of generators of an algebra. As a consequence, the solution
of the Schr\"odinger equation then becomes an algebraic problem, which can be
attacked using the tools of group theory. In addition to providing a general
approach to solving the problem of interest, the SGA also provides as special
cases, exactly solvable limits of the  models. These are known as
dynamical  symmetries, and can often be found by suitable
choices of the parameters of a model. Typically one has more
forms of the Hamiltonian which are exactly
solvable, than dynamical symmetries predicted from group theory\cite{ds}.

The dynamical symmetry problems up to now have focussed essentially on the
description of bound and scattering states. By mapping the quantum
mechanical problem onto an algebraic structure,  representation
theory  immediately provides the excitation energies and
eigenstates. However, there is one missing category in  the group
theoretical description of quantum
systems, namely,  Hamiltonians with periodic potentials. In
this case, the energy spectrum is characterized by energy bands
and gaps. We extend the group
theoretical  framework by showing that dynamical
symmetry and SGA techniques can be applied to such problems, where dispersion
relations, transfer matrices and energy bands can be understood in the context
of representation theory. 

Consider Hamiltonians in one dimension of the form:
\begin{equation}
H=-\frac{\hbar ^2}{2M}\frac{d^2}{d\rho ^2}+V(\rho ).
\end{equation}
For the P\"oschl-Teller potential, $V(\rho )\propto 1/\cosh
^2(\rho /\rho _0)$, it is well known that bound states can be
expressed as discrete representations of $SU(2)$, and scattering
states by the principal series of $SU(1,1)$, for which
$j=-1/2+ik$, $k>0$\cite{fia,wo,fib}.  In passing from $SU(2)$ (
where $[J_z,J_\pm]=\pm J_\pm$ and $[J_+,J_-]=2J_z$) to $SU(1,1)$
(obtained by changing the sign in $[J_+,J_-]=-2J_z$), the
Casimir invariant, $J^2$ changes from $J^2=J_x^2+J_y^2+J_z^2$
(sphere) to $J^2=J_z^2-J_x^2-J_y^2$ (hyperboloid), and the
representation quantum numbers $j,m$ fall into three well known
series: the principal, discrete and complementary.
 However, if one has a periodic
potential $V(\rho )=V(\rho +a)$, for some constant $a$, it is
not at all clear how this fits into the scheme provided by
representation theory. In this case we would like to derive the
transfer matrices and dispersion relations.

When one considers the eigenstates in an energy band, Bloch's
theorem states that it must have the form
$\Psi_k(\rho)=\phi_k(\rho)\exp(ik\rho)$, where $\phi_k(\rho+a)=\phi_k(\rho)$
has the periodicity of the potential.  Group theory typically
provides single-valued wavefunctions, 
suitable for band edges.  In the $SU(1,1)$
dynamical symmetries we construct, we will see that
wavefunctions can be obtained by {\it (a)}  using multiple-valued
projective representations $(j,m)$ (labeled by $J^2$ and $J_z$
and associated with equivalence classes of vectors defined up to
a phase, which is natural in Hilbert space),
and {\it (b)} by using a basis 
$\mid j,\lambda\rangle$ labeled by $J^2$ and a non-compact
generator of $SU(1,1)$ such as $J_x$. 
Several surprising aspects will emerge, such as
the use of the complementary series and the need for non-unitary
representations.

In order to see how dynamical symmetry methods and
representation theory might apply to Hamiltonian with band
structure, we must first construct a dynamical symmetry for a
periodic potential. It is possible to do so with the Scarf
potential\cite{sc}: $V(\rho )\propto 1/\sin ^2(\rho /\rho _0)$.
In dimensionless variables:
\begin{equation}\label{sca}
H_{Sc}\Psi (x)=\left[ -\frac{d^2}{dx^2}+\frac A{\sin ^2x}\right] \Psi (x)=
{\cal E}\Psi (x).
\end{equation}
(The energy is now $E=(\hbar^2/2M\rho_0^2){\cal E}$.)  
This potential is known to have band structure 
for $-\frac14<A<0$\cite{sc}. By suitable choices of
generators, we can realize $H_{Sc}$ as a Casimir
invariant. There are different ways to do this, but not all are
fruitful.  For instance, in $SU(2)$
we obtain $A=m^2-1/4$, and ${\cal E}=(j+1/2)^2$ . Here $H_{Sc}={\bf J}^2+1/4$ 
is simply related to the Casimir invariant ${\bf J}^2$
of $SU(2)$. This realization is unsuitable since {\it (i)} the
discrete representations give only periodic solutions and {\it
(ii)} the coupling $A$ is never in the range of physical
interest.

If we choose the generators of $SU(1,1)$ in the form
$J_{\pm }=e^{\pm i\phi }(\mp \sin\theta \partial _\theta +i\cos \theta 
\partial _\phi )$, and $J_z=-i\partial _\phi$, we can realize $H_{Sc}$
with $A=j(j+1)$ and  ${\cal E}=m^2$:
\begin{equation}
H_{Sc}\Psi _{jm}=J_z^2\Psi _{jm}=m^2\Psi _{jm}.  \label{scarf}
\end{equation}
This is more natural, since a given strength of the potential corresponds to a
fixed representation $j$, and the eigenvalue is related to the projection $m$.
(We have replaced $\theta $ by $x$ in relating Eq. \ref{scarf} for Eq. 
\ref{sca}).  To describe band spectra in $SU(1,1)$, we pass to
the projective unitary representations\cite{su}, which have 
three classes: the discrete
series $D_j^{\pm }$, with $j<0$, the principal series where $j=-1/2+i\sigma $,
($\sigma>0 $), and the complementary series with $-1/2<j<0.$ 
 While the discrete and principal series of $SU(1,1)$ has
found application to problems such as the P\"oschl-Teller
potential discussed above,  the remaining complementary series has
found little or no application in physics. Surprisingly, this series
is precisely what is needed for band structure, and is essential to complete
the dynamical symmetry approach to quantum systems.

The complementary series is shown in Fig. 1 (top) and corresponds to potentials
with $-1/4<A<0$, exactly the region of interest. The
unitary representations are labeled by two quantum numbers $j$,$m$, with
$m=m_0\pm n$ ($n=0,1,...$), where $0\leq m_0<1$ and $ m_0(m_0-1)>j(j+1)\geq
-\frac 14$. In the figure, we plot a generic line $j(j+1)=const$ (dots) 
together with the parabola $m_0(m_0-1)$ (solid). We also show the periodic
repetition of this parabola given by $(m\pm n)(m\pm n-1)$
(dashes). The values of $j,m$
which correspond to unitary representations are then found to be $n-j\leq m\leq
j+n+1$ as well as $-(j+n+1)\leq m\leq -(n-j)$. The relation of these
representations to the band structure is shown in Fig. 1(bottom). There we plot
the eigenvalue $m^2$ as a function of $j$, related to the strength of the 
potential through $A=j(j+1)$. The range of
$m$ corresponding to unitary representations give the {\it energy gaps}. The
{\it non-unitary} representations correspond to the {\it energy bands} (shaded
region). We know that the non-unitary representations must correspond to the
energy bands from an analysis of the $j\rightarrow 0$ limit. 
Here the strength of the
potential $A=0$, and we have free motion, where all values of $m$ must be
allowed, which from Fig. 1(top) corresponds to the non-unitary representations.

The discrete (projective) representations are readily understood
as well. The series $D_j^{\pm}$ provides the lower band edges, as
shown by the boxes in Fig. 1(bottom) for a particular choice of potential given
by $j$ (dashes). (Both series $\pm $ are degenerate and give the same
spectrum.) The upper band edges are obtained by realizing that the strength
$A=j(j+1)$ of the potential is invariant under the  transformation
$j\rightarrow -(j+1)$. The discrete series $D_{-1-j}^{\pm }$ then corresponds
to the upper band edges (circles). The principal series 
has a potential with strength $A< -\frac 14$, for which
the Hamiltonian is no longer self-adjoint, and is of no physical
interest. 

The transfer matrix $T$ can be computed directly from the $SU(1,1)$
wavefunctions (where care is taken at the singularity\cite{lk}):
\begin{eqnarray}
T &=&\left( 
\begin{array}{cc}
\alpha  & \beta  \\ 
\beta ^{*} & \alpha ^{*}
\end{array}
\right)  \\
\alpha  &=&e^{-i\pi m}\left[ \frac{\cos \pi m}{\sin \pi (j+\frac 12)}
-i\left( \frac 2m\frac{\Gamma \left( \frac{1-j+m}2\right) \Gamma \left( 
\frac{1-m-j}2\right) }{\Gamma \left( \frac{m-j}2\right) \Gamma \left(-\frac{
j+m}2\right) }\right. \right. \nonumber \\
&-&\frac m2\left. \frac{\Gamma \left( \frac{m+j+1}2\right) \Gamma \left( 
\frac{1-m+j}2\right) }{\Gamma \left( \frac{2+m+j}2\right) \Gamma \left( \frac{
2+j-m}2\right) }\right) \left. \frac{\cos \pi \frac{j+m}{2}\cos\pi\frac{j-m}{2}}
{\sin\pi(j+\frac{1}{2})}
\right]  \nonumber\\ 
\beta  &=& ie^{i\pi m}\left( \frac m2\frac{\Gamma \left( \frac{1+j+m}2
\right) \Gamma \left( \frac{1+j-m}2\right) }{\Gamma \left( \frac{2+j+m}2
\right) \Gamma \left( \frac{2+j-m}2\right) }\right.  \nonumber\\
&+&\frac 2m\left. \frac{\Gamma \left( \frac{1-j+m}2\right) \Gamma \left( 
\frac{1-m-j}2\right) }{\Gamma \left(-\frac{j-m}2\right) \Gamma \left( -
\frac{j+m}2\right) }\right) \frac{\cos \pi\frac{j-m}2\cos\pi\frac{j+m}{2} }
{\sin\pi (j+\frac{1}{2})}. \nonumber
\end{eqnarray}
Taking $\cos k\pi =Re(\alpha e^{i\pi m})$, we obtain the dispersion relation 
$\cos \pi k=\cos \pi m/\sin \pi (j+\frac 12)$, or
\begin{equation}
{\cal E}(k)=\frac{1}{\pi^2}\left[\cos^{-1}\left(\cos \pi k \sin \pi
  \sqrt{A+\frac{1}{4}}\right)\right]^2,
\end{equation}
which agrees with that derived by Scarf\cite{sc}.
The bands are restricted to the non-unitary parts of the complementary
series of $SU(1,1)$.

Using the properties of our Scarf dynamical symmetry, we can consider a
generalization of this potential in $SO(2,2)\approx SU(1,1)\times SU(1,1)$. The
states are labeled by $j,m,j',c$ through the direct product $\left|
jm\right\rangle \times \left| j'c\right\rangle $ of $SU(1,1)$
states.  It is possible to construct 
generators of the $SU(1,1)$ algebras, denoted ${\bf J}$ and ${\bf
  K}$, for which $J^2=K^2$, so that
$j=j'$, and which has the
following periodic Hamiltonian as dynamical symmetry: 
\begin{eqnarray}
\left[ -\frac{d^2}{dx^2}\right. &+&\left.
        \frac{(m+c)^2-\frac 14}{\sin ^2x}\right.
        +\left. \frac{4(j+\frac 12)^2-\frac 14}{\cos ^2x}\right] \Psi _{jmc} 
\nonumber \\
 &=&(J_z-K_z)^2\Psi _{jmc}=(m-c)^2\Psi _{jmc}.  \label{escarf}
\end{eqnarray}
(This can also be expressed as $V(x)=(a+b\cos x)/\sin^2 x$.) 
We consider the band structure for  $\left| m+c\right| \leq \frac 12$ 
and $-\frac 34<j<-\frac 14$. Again by symmetry, we
use the regime  $-\frac 12\leq j<-\frac 14$. 
To determine the band structure, we take the projective representations of 
$SO(2,2)$ obtained through the direct product of the results of Fig. 1 (top).
These are shown in Fig. 2, where we plot the complementary series for 
$SO(2,2)$ as
a function of $m-c$ and $m+c$ for three values of $j.$ The shaded areas 
correspond to unitary representations. The striped regions are obtained from
direct products of unitary representations, while the checkered regions are
direct products of non-unitary representations, which can be made unitary in
the strip  $\left| m+c\right| \leq \frac 12.$ The representations in
Fig. 2 correspond to {\it (a)} $j=-0.45$,{\it \ (b)} $j=-0.35$ and{\it \ (c)}
$j=-0.25$. The energy gaps are indicated by the shaded
regions. To obtain the bands for a given potential ($m+c$ and $j$ constant), 
one chooses $j$ and the value of $m+c$ in the range $[-1/2,1/2]$, and
draws the vertical line. The values of $m-c$ which are in the unshaded
regions, including the boundaries, correspond to the energy bands.  One can see
that as $j\rightarrow -\frac 12$, the bands vanish, leaving only discrete
eigenvalues, analogous to Fig. 1(bottom). 

The transfer matrix can also be derived (which we omit here for brevity),
and results in the dispersion relation: 
\begin{equation}
\cos (\pi k)=\frac{\cos (2j+1)\pi \cos (m+c)\pi +\cos (m-c)\pi }{\sin
(2j+1)\pi \sin (m+c)\pi }.
\end{equation}
This $SO(2,2)$ Hamiltonian has three limits where it reduces
to the Scarf
potential: {\it (a)} $j=-\frac 14$, {\it (b)} $m+c=\pm \frac 12$, and {\it 
(c) }$m+c=2j+1.$ In these cases, the range of $m-c$ agrees with
that in Fig. 1 (with twice the period for case {\it (c)} ), and the dispersion
relation agrees with the Scarf result.

We have realized the above dynamical symmetry band problems using projective
representations of $SU(1,1)$ in the $J^2,J_z$ basis. There are other ways in
which one can realize band structure as well, using bases defined through
non-compact generators, such as $J^2,J_x$.  In this case the states $\mid
j,\lambda\rangle$ are continuous, with $\lambda$ a real number. The final case
we would like to discuss here is a more general SGA case of  this type.
Consider the 1-d crystal formed from the superposition of P\"oschl-Teller
potentials $V(\rho)=-(j(j+1)/\rho _0^2)\sum_n\cosh ^{-2}([\rho -na]/\rho _0)$.
Through the linear transformation $ \rho =(i\pi\rho_0/2K)x - (a+i\pi\rho_0)/2$,
where $K$ and $K'$ are elliptic integrals, the  Hamiltonian reduces to
the Lam\'e  equation, which has a periodic potential of real period
$2K$\cite{suth,ww}
:
\begin{equation}
\left[ -\frac{d^2}{dx^2}+j(j+1)\kappa ^2{\rm sn}^2(x|\kappa )\right] \Psi
     ={\cal E}\Psi .
\end{equation}
Here sn$(x|\kappa )=$sn$(x)$ is the doubly periodic Jacobi elliptic
function of  modulus $\kappa$. 
This equation can be related to the non-dynamical symmetry
problem\cite{fia,miller}
\begin{equation}
H=J_x^2+\kappa ^2J_y^2.
\end{equation}
When the generators ${\bf J}$ are written in conical coordinates
$(\theta,\phi)$, the above
Hamiltonian decouples into two Lam\'e equations, one in $\theta$
and the other in $\phi$. The band edges are obtained from $SU(2)$\cite{fia},
which should be contrasted to the previous examples where the edges were
classified according to the non-compact algebras $SU(1,1)$ and $SO(2,2)$. 
For integer $j$, there are $j+1$ energy bands.

Non-periodic solutions of Lam\'e's equation can be expressed in terms of theta
functions\cite{ww}. We can explicitly construct
the transfer matrix and the dispersion relation for the Lam\'e equation with
integer $j$:
\begin{equation}
\alpha =\cos 2k({\cal E})K\quad \beta =i\frac{\sin 2k({\cal E})K}{\sum_{n=1}^j
\mathop{\rm sn}\alpha _n\mathop{\rm dn}\alpha _n/\mathop{\rm cn}\alpha _n}.
\end{equation}
where the dispersion relation $k(E)$ is expressed in terms of the Jacobi zeta 
function $Z$ of modulus $\kappa$:
\begin{equation}
k({\cal E})=-i\sum_{n=1}^j Z(\alpha _n\mid \kappa^2)+j\frac \pi {2K}.
\end{equation}
The parameters $\alpha _n$ are solutions of $j$ coupled equations which
depend upon the eigenvalue ${\cal E}$\cite{ww}: 
\begin{eqnarray}
{\cal E} &=&\sum_{n=1}^j\mathop{\rm ns}^2\alpha _n-\left[ \sum_{n=1}^j
         \mathop{\rm cn}\alpha _n\mathop{\rm ds}\alpha _n\right] ^2 \\
0 &=&\sum_{p=1}^j\frac{
   \mathop{\rm sn}\alpha _p\mathop{\rm dn}\alpha_p\mathop{\rm cn}\alpha_p 
   - \mathop{\rm sn}\alpha_n\mathop{\rm dn}\alpha_n \mathop{\rm cn}\alpha_n}
   { \mathop{\rm sn}^2\alpha _p-\mathop{\rm sn}^2\alpha _n}
         \quad (p\neq n).\nonumber
\end{eqnarray}
This transfer matrix has the following properties. {\it (a)} In the limit
$\kappa =1$, the Hamiltonian becomes the  P\"oschl-Teller potential, which is
not periodic, and the denominator of $\beta$ becomes $i\sqrt{{\cal
E}-j(j+1)}=ik$. Here we obtain the transfer matrix with matrix elements
$\alpha_{PT} = \Gamma (ik)\Gamma (1+ik)/ \Gamma (1+ik+j)\Gamma (ik-j)$ and
$\beta_{PT} \rightarrow 0$, the desired result for integer $j$.  {\it (b)} In
the limit $\kappa =0$, the Hamiltonian (9) becomes $H=J_x^2$ while (8) is free
motion. In the basis $\mid j,\lambda\rangle$, (9) has the
desired free particle dispersion, ${\cal E}=\lambda^2\geq 0$. We also
recover the free particle transfer  matrix. {\it (c)} There are additional
connections one can make with the Mathieu equation when $\kappa \rightarrow 0,$
$j\rightarrow \infty $, and $\kappa ^2j(j+1)=const.$, as well as connections to
the Scarf potential. Because the equations (12) must generally be solved
numerically, we show the type of result one finds for
$j=1$. Here there is one root $\alpha$, and the wavefunction in Bloch form is:
\begin{equation}
\Psi_k(x)=\left[ \frac{H(x+\alpha )}{\Theta (x)}\exp \left( \frac{i\pi x}{2K
}\right) \right] \exp (-ikx),
\end{equation}
where $H$ and $\Theta$ are theta functions, and $\alpha$ is determined 
from
${\cal E}=\mathop{\rm dn}^2\alpha +\kappa ^2$, which restricts
the energy to the bands $\kappa^2\leq {\cal E}\leq 1$ and 
$1+\kappa ^2\leq {\cal E}<\infty .$ For the lower band, 
$\alpha=K+i\beta$ where $\beta$ ranges from $K'$ to 0, while for
the upper band $\alpha=i\beta$, where $\beta$ ranges from $0$ to
$K'$. The dispersion relation then has the form
\begin{equation}
k({\cal E}) = -Z(\beta |1-\kappa ^2)+\frac \pi {2K}\left( 1-\frac \beta {
                  K^{\prime }}\right) + {\cal E}',
\end{equation}
where ${\cal E}'= \sqrt{(1-{\cal E})({\cal E}-\kappa ^2)/
(1+\kappa ^2-{\cal E})}$ for the lower band, and  ${\cal E}' = 
\sqrt{({\cal E}-\kappa^2)({\cal E}-1-\kappa ^2)/({\cal E}-1)}$
for the upper band.
In contrast to the Scarf problem, we see here that the band edges are given by
the discrete representations of $SU(2)$. The bands correspond to a
diagonalization in the $SU(1,1)$ basis, such as that defined by $J$ and $J_x$. 
This is evident if we examine the limit $\kappa =0$, where we expect a
continuous spectrum.

We have shown that SGA and dynamical symmetry techniques can be
applied to Hamiltonians with periodic potentials, and band
structure can arise naturally from representation theory.  This
fills a long-standing gap in the algebraic approach to quantum systems.
Hamiltonians such as the Scarf's are now reduced to problems as
simple as solving $H=J_z^2$, using representations with
multi-valued wavefunctions. The complementary series
now finds natural application in the band structure problem,
where its non-unitary representations are important. Further,
both compact and non-compact Lie algebras play a role in the
band edges - $SU(1,1)$ for the Scarf and extended Scarf
potentials, and $SU(2)$ for the Lam\'e equation. We expect that
dynamical symmetries exist for potentials in higher dimensions,
where the point group of the lattice will play a role in the
representations of algebras such as $SU(n,m)$. Within these
representations one should find the complete spectral solutions to the
problems.

We would like to thank F. Iachello for many useful discussions. This work
was supported by DOE grant DE-FG02-91ER40608.

\begin{figure}
\epsfxsize=8 cm
\centerline{\epsffile{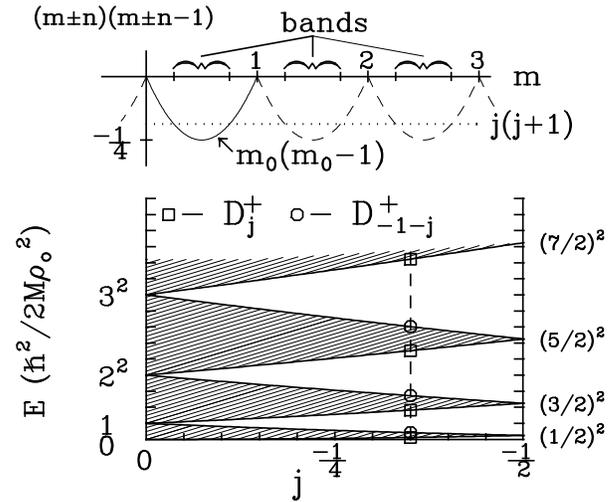}}
\vspace{2 mm}
\caption{(Top) Complementary series of SU(1,1) labeled by
$j,m$. The unitary (projective) representations satisfy
$m_0(m_0-1)>j(j+1)$, where $m=m_0\pm n$ ($n=0,1,...)$ and $0\leq
m_0< 1$, and coincides with the energy gaps.
(Bottom) The corresponding Scarf energy spectrum
$E=[\hbar^2/(2M\rho_0^2)]m^2$ is plotted as a function of
$j$. The band edges are given by the discrete series and the
bands (shaded) by the non-unitary part of the complementary
series.}
\label{fig1}
\end{figure}

\begin{figure}
\epsfxsize=8 cm
\centerline{\epsffile{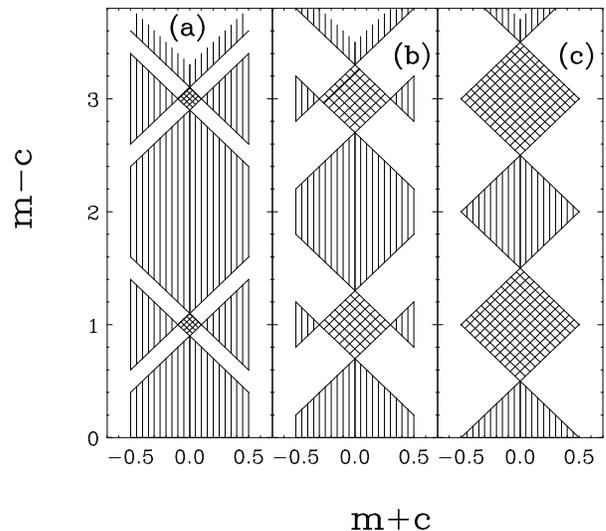}}
\vspace{2 mm}
\caption{Unitary representations (shaded regions) of $SO(2,2)$
obtained from Fig. 1 (top), restricted to the physical region
$|m+c|\leq 1/2$, for selected values $(a)$ $j=-0.45$, $(b)$
$j=-0.35$ and $(c)$ $j=-0.25$. The physical spectrum, given by
$(m-c)^2$, is obtained by choosing a value of $m+c$ and $j$
(i.e. a specific potential). The bands are the values of $m-c$
which are non-unitary (in the unshaded regions).}
\label{fig2}
\end{figure}

\end{document}